\pdfoutput=1
\documentclass[a4paper,12pt,english]{article}

\usepackage[pdftex]{graphicx}
\usepackage{graphicx}
\usepackage{colordvi}
\usepackage{fancybox}
\usepackage{cancel}
\usepackage{amsmath}
\usepackage{amssymb}
\usepackage{caption}
\usepackage{appendix}
\usepackage{hyperref} %naredi dokument aktiven (v DVI in PDF)
\usepackage{multirow}
\hypersetup{colorlinks=true, linkcolor=black, urlcolor=blue}
\usepackage{float}
\usepackage{cite}
\usepackage{mathtools}

\usepackage{wasysym}

\usepackage{color}
\definecolor{gray}{rgb}{0.5,0.5,0.5}

\newcommand\lsim{\mathrel{\rlap{\lower4pt\hbox{\hskip1pt$\sim$}}
    \raise1pt\hbox{$<$}}}
\newcommand\gsim{\mathrel{\rlap{\lower4pt\hbox{\hskip1pt$\sim$}}
    \raise1pt\hbox{$>$}}}

\newcommand{\beq}{\begin{equation}}
\newcommand{\eeq}{\end{equation}}
\newcommand{\bea}{\begin{eqnarray}}
\newcommand{\eea}{\end{eqnarray}}
\newcommand{\bem}{\begin{pmatrix}}
\newcommand{\eem}{\end{pmatrix}}
\newcommand{\noi}{\noindent}
\newcommand{\non}{\nonumber}

\newcommand{\bet}{\begin{itemize}}
\newcommand{\eet}{\end{itemize}}
\newcommand{\ben}{\begin{enumerate}}
\newcommand{\een}{\end{enumerate}}

\begin{document}
\sloppy

\numberwithin{equation}{section}

\begin{flushright}
%MITP-preprint number ?
%December, 2015
\end{flushright}

\bigskip

\begin{center}

{\Large\bf Kutasov-Seiberg dualities and cyclotomic polynomials}
%{\Large\bf Classification of Seiberg dualities from number theory}
\vspace{1cm}

\centerline{Borut Bajc$^{a,}$\footnote{borut.bajc@ijs.si}} 

\vspace{0.5cm}
\centerline{$^{a}$ {\it\small J.\ Stefan Institute, 1000 Ljubljana, Slovenia}}
%\centerline{Steven Abel$^{a,}$\footnote{s.a.abel@durham.ac.uk}, Borut Bajc$^{b,}$\footnote{borut.bajc@ijs.si} and 
%Francesco Sannino$^{c,}$\footnote{sannino@cp3.sdu.dk}}
%
%\vspace{0.5cm}
%\centerline{$^{a}$ {\it\small IPPP, Durham University, South Road, Durham, DH1 3LE}}
%\centerline{$^{b}$ {\it\small J.\ Stefan Institute, 1000 Ljubljana, Slovenia}}
%\centerline{$^{b}$ {\it\small CP$^3$-Origins \& the Danish IAS, University of Southern Denmark,  Denmark}}
%%\centerline{{\it\small }}
\end{center}

\bigskip

\begin{abstract}
We classify all Kutasov-Seiberg type dualities in large $N_c$ SQCD with adjoints of rational $R$-charges. 
This is done by equating 
the superconformal index of the electric and magnetic theories: the obtained equation has a solution each time 
some product of cyclotomic polynomials has only positive coefficients. In this way we easily reproduce without any 
reference to the superpotential or the choice of the equations of motion (classical chiral ring) all the known 
dualities from the literature, while adding to them a new family with two adjoints with $R$ charges $\frac{2}{2k+1}$ 
and $\frac{2(k+1)}{2k+1}$ for all integers $k>1$. We argue that these new fixed points could be in their appropriate 
conformal windows and in some range of the Yukawas involved a low energy limit of the $D_{2k+2}$ fixed point. 
We try to clarify some issues connected to the difference between classical and quantum chiral ring of this new solution. 
\end{abstract}

\clearpage

\tableofcontents
% \newpage

\section{Introduction}

There are many known candidates for IR fixed points in SQCD with $N_A$ adjoints connected with a free theory in the UV: they are 
described for no ($N_A=0$) adjoints in \cite{Seiberg:1994pq}, for $N_A=1$ adjoint in \cite{Kutasov:1995ve,Kutasov:1995np,Kutasov:1995ss}, 
and $N_A=2$ adjoints in \cite{Intriligator:2003mi}. Larger values of $N_A>2$ are unavailable for branches of the flows which end 
up eventually into a free UV fixed point. Several of the above fixed points have, in a specified conformal window, a known candidate 
for a non-trivial dual: \cite{Seiberg:1994pq} in the case $N_A=0$, \cite{Kutasov:2003iy} for $N_A=1$, while for  $N_A=2$ the known 
duals are for $D_{k+2}$ \cite{Brodie:1996vx,Brodie:1996xm,Intriligator:2003mi} and $E_7$ 
\cite{Kutasov:2014yqa,Kutasov:2014wwa}\footnote{Particular aspects of dualities in SQCD with one and two adjoint have been discussed in 
\cite{Mazzucato:2004mq}, \cite{Mazzucato:2005fe}.}. 

In this paper we will try to derive in a different way these fixed points of SQCD with adjoints with known duals. 
We will assume that the $R$-charges of the adjoints are rational numbers, as if determined from a marginal 
superpotential. We will show that with such an assumption a complete (although in principle infinite) classification of all 
solutions with valid duality through the equality of the superconformal index 
\cite{Romelsberger:2005eg,Kinney:2005ej,Romelsberger:2007ec} at large $N_c$ \cite{Dolan:2008qi} can be found. This is because, 
as we will see, such an equality can be written as a factorisation of a polynomial $y^n-1$ for integer $n>1$ into a product of a 
polynomial with only non-negative integer coefficients (representing the contribution of the mesons) with a specific (antipalindromic) 
polynomial (representing the contributions of all the gauge adjoints of the theory) to be specified later on. It 
is well known that every polynomial of the form $y^n-1$ for a fixed integer $n$ can be written as a unique product of so-called 
cyclotomic polynomials \cite{cyclotomic}. Our classification of all duals is thus transformed into a classification of all products of distinct 
cyclotomic polynomials with only non-negative coefficients. Although we were unable to find an explicit analytic classification 
of this mathematical problem, the formulation helps in finding the solutions: choose $n>1$, find the factorisation of $y^n-1$ 
into product of cyclotomic polynomials, and find all the partial products which end up into polynomials with non-negative 
coefficients.

With such a recipe we can easily reproduce the known dualities in the case of $N_A\leq2$, i.e. SQCD, $A_k$, $D_{k+2}$ 
and $E_7$. We can however find out other possible rational solutions for $R$-charges between $0$ and $2$, which we were unable 
to locate in the existing literature. We will mention here the family with $N_A=2$ and the adjoint $R$-charges 
$\frac{2}{2k+1}$ and $\frac{2(k+1)}{2k+1}$ for integer $k>1$. Although it is easy to write down a superpotential which, when 
marginal, produces such $R$-charges, the classical determination of the mesons needed in the dual theory 
using the equations of motion seems incomplete. On the other side using the equality of the superconformal indices the mesons' 
$R$-charges and thus their structure come out automatically as an output. We will show how this comes about in the 
quantum chiral ring from the explicit evaluation of few low order terms of the superconformal index. We found also a possible 
interpretations for such new fixed points: they could be IR limits of the $D_{2k+2}$ fixed point, perturbed by a relevant operator 
in a specified conformal window. 

We will first give in section \ref{index} a short summary from the literature of the calculation of the superconformal index and its 
use in comparing dual theories at large $N_c$. We will use it to derive the main equation, i.e. eq. (\ref{primerjava}) for SQCD 
with $N_A$ adjoints. In the following section \ref{klasifikacija} we will then show how to go through the ordered but infinite solutions 
of the main equation (\ref{primerjava}), by using the notion of cyclotomic polynomials. Here we will also describe some old and 
new solutions from our new way of classifying them. In section \ref{nova} we will take some time to interpret the new solution found, 
i.e. in which flow it may appear, and check its quantum chiral ring from the superconformal index. We will then conclude in section 
\ref{conclusion} and add three appendices.

\section{\label{index}The superconformal index}

We first summarise the computation of the superconformal index (for a nice review see for example 
\cite{Rastelli:2016tbz}) following mainly \cite{Romelsberger:2007ec,Dolan:2008qi,Spiridonov:2009za}. 

The first step is to calculate the index over the single particle states through

\beq
\label{indeks}
i(t,x,z,y)=\frac{2t^2-t(x+x^{-1})}{(1-tx)(1-tx^{-1})}\chi_{adj}(z)
+\sum_j\frac{t^{R_J}\chi_{F(j)}(y)\chi_{G(j)}(z)-t^{2-R_j}\chi_{\bar F(j)}(y)\chi_{\bar G(j)}(z)}
{(1-tx)(1-tx^{-1})}
\eeq

\noi
where $t$, $x$, $z$, $y$ are fugacities (chemical potentials) defined for the generators of conserved 
quantities in the theory considered: $t$ and $x$ for the superconformal group on $R\times S^3$ where
 the index is defined, while $z$ and $y$ are for the gauge and global symmetry groups, respectively. 
 $\chi$ is the group character which is in SU(N)

\bea
\chi_{fundamental}(u)= p_{N}(u)\equiv\sum_{i=1}^Nu_i&,&\chi_{antifundamental}(u)=p_{N}(u^{-1})\\
\chi_{adjoint}(u)=p_{N}(u)p_{N}(u^{-1})-1&,&\chi_{singlet}=1
\eea

\noi
and $\prod_{i=1}^{N}u_i=1$.

In (\ref{indeks}) the first term is the contribution of the gauge fields, while $j$ in the sum of the second term 
runs over all the chiral superfields: the term proportional to $t^{R_j}$ is the contribution of the boson $\phi_j$ 
with $R$-charge $R_j$ while the term proportional to $t^{2-R_j}$ is the contribution of the fermion $\bar\psi_j$. 
Notice that $\psi_j$ is the superpartner of $\phi_j$. $F(j)$ and $G(j)$ stand for the representations of the flavour 
and gauge group respectively of the chiral superfield $j$.

Now it is easy to write the index for our specific case of $N_A$ adjoints with R charges $R_i$ and 
$N_f$ pairs of vectorlike quarks $Q+\tilde Q$:

\bea
\label{indekszanas}
i(t,x,y,\tilde y,v,z)&=&\frac{2t^2-t(x+x^{-1})+\sum_{a=1}^{N_A}(t^{R_a}-t^{2-R_a})}{(1-tx)(1-tx^{-1})}
\left(p_N(z)p_N(z^{-1})-1\right)\non\\
&+&\frac{\left(t^{R_Q}p_{N_f}(y)-t^{2-R_Q}p_{N_f}(\tilde y)\right)v}
{(1-tx)(1-tx^{-1})}p_N(z)\non\\
&+&\frac{\left(t^{R_Q}p_{N_f}(\tilde y^{-1})-t^{2-R_Q}p_{N_f}(y^{-1})\right)v^{-1}}
{(1-tx)(1-tx^{-1})}p_N(z^{-1})
\eea

\noi
where instead of a single $y$ we used $y$ itself for the group SU($N_f$)$_Q$, $\tilde y$ for 
SU($N_f$)$_{\tilde Q}$, and $v$ for the baryonic U(1)$_B$.

In short we can thus write

\beq
i(w,z)=f(w)\left(p_N(z)p_N(z^{-1})-1\right)+g(w)p_N(z)
+\bar g(w)p_N(z^{-1})+h(w)
\eeq

\noi
where we denoted by $w$ all the fugacities except the gauge $z$ (i.e. in our case $t,x,y,\tilde y,v$). 

For the electric theory we are considering 

\bea
\label{fE}
f_E&=&\frac{2t^2-t(x+x^{-1})+\sum_{a=1}^{N_A}(t^{R_a}-t^{2-R_a})}{(1-tx)(1-tx^{-1})}\\
\label{gE}
g_E&=&\frac{\left(t^{R_Q}p_{N_f}(y)-t^{2-R_Q}p_{N_f}(\tilde y)\right)v}{(1-tx)(1-tx^{-1})}\\
\label{gEbar}
\bar g_E&=&\frac{\left(t^{R_Q}p_{N_f}(\tilde y^{-1})-t^{2-R_Q}p_{N_f}(y^{-1})\right)v^{-1}}{(1-tx)(1-tx^{-1})}\\
h_E&=&0
\eea

\noi
while the magnetic theory of Seiberg type has

\bea
f_M&=&\frac{2t^2-t(x+x^{-1})+\sum_{a=1}^{N_A}(t^{R_a}-t^{2-R_a})}{(1-tx)(1-tx^{-1})}\\
g_M&=&\frac{\left(t^{R_q}p_{N_f}(y^{-1})-t^{2-R_q}p_{N_f}(\tilde y^{-1})\right)v^{N_c/\tilde N_c}}{(1-tx)(1-tx^{-1})}\\
\bar g_M&=&\frac{\left(t^{R_q}p_{N_f}(\tilde y)-t^{2-R_q}p_{N_f}(y)\right)v^{-N_c/\tilde N_c}}{(1-tx)(1-tx^{-1})}\\
\label{hM}
h_M&=&\frac{\sum_j\left(t^{2R_Q+R_j}p_{N_f}(y)p_{N_f}(\tilde y^{-1})-
t^{2-2R_Q-R_j}p_{N_f}(y^{-1})p_{N_f}(\tilde y)\right)}{(1-tx)(1-tx^{-1})}
\eea

\noi
where in the expression for $h_M$ the index $j$ runs over the mesons, and $R_j+2R_Q$ are their $R$-charges.
Notice that $f_E=f_M$.

The full superconformal index is finally the gauge invariant part of the Plethystic exponential of (\ref{indeks})

\beq
\label{celindeks}
{\cal I}(w)=\int_Gd\mu\exp\left(\sum_{i=1}^\infty\frac{1}{n}i(w^n,z^n)\right)
\eeq

\noi
where we integrate over the gauge group $G$ and $w$ stands for all the non-gauge fugacities.

For simplicity we will here consider the large $N$ limit, when the full index (\ref{celindeks}) can be reduced 
\cite{Dolan:2008qi} to

\beq
\label{gaugeinvariantindex}
{\cal I}(w)=\exp\left(\sum_{n=1}^\infty\frac{1}{n}\left(\frac{g(w^n)\bar g(w^n)}{1-f(w^n)}-f(w^n)+h(w^n)\right)\right)\prod_{n=1}^\infty
\frac{1}{1-f(w^n)}
\eeq

The electric and magnetic theory should have the same superconformal index, so in the large $N$ limit 
we need to satisfy

\beq
\frac{g_E\bar g_E-g_M\bar g_M}{1-f}=h_M-h_E
\eeq

It is not difficult to show that this boils down to the relation

\beq
\label{primerjava}
\sum_{j=1}^\alpha t^{R_j}=\frac{t^{2\alpha\left(\sum_{a=1}^{N_A}(R_a-1)+1\right)}-1}
{t^2-1+\sum_{a=1}^{N_A}\left(t^{R_a}-t^{2-R_a}\right)}
\eeq

\noi
where $\alpha$ is equal to the number of independent mesons in the chiral ring and 

\beq
\label{Nctilde}
\tilde N_c=\alpha N_f-N_c
\eeq

The strategy is thus \cite{Kutasov:2014wwa} to get such $R_a$, $R_j$, $N_A$, $\alpha$, for which 
(\ref{primerjava}) is satisfied. There are some general arguments on how the solutions of (\ref{primerjava}) 
must look like \cite{Kutasov:2014wwa}. For example, the absolute value of all zeros of the denominator 
must be unity, since only such zeros are in the numerator, or that each zero of the denominator must be 
unique, since there is no double zero in the numerator \cite{Kutasov:2014wwa}. 

In theories in which we have a classical truncation of the chiral ring (this means that the chiral ring 
is finite - i.e. its dimension is independent of $N_c$ - directly following the equations of motion), one 
can explicitly find $\alpha$ and $R_j$, while $N_A$ is known from the definition of the theory and 
$R_a$ from the superpotential. This is known to work well in the $A_k$ ($k$ integer positive) theory 
($N_A=2, \alpha=k$)

\beq
W_{A_k}=Tr\left(X^{k+1}+Y^2\right)
\eeq

\noi
(SQCD with 1 adjoint) and $D_{k+2}$ ($k$ odd positive integer) with 
($N_A=2,\alpha=3k$)

\beq
W_{D_{k+2}}=Tr\left(X^{k+1}+XY^2\right)
\eeq

\noi
according to the classification of \cite{Intriligator:2003mi}. In such cases equation (\ref{primerjava}) is 
just an extra check of the duality, but we do not learn more than with usual techniques. 

The power of the superconformal index is in cases in which there is no classical truncation of the chiral ring, 
for example in $D_{k+2}$ for even $k$ or $E_7$ ($N_A=2, \alpha=30$):

\beq
W_{E_7}=Tr\left(Y^3+YX^3\right)
\eeq

\noi
where the above computation suggests that $\alpha=30$ \cite{Kutasov:2014wwa}. 

The same technique proves \cite{Kutasov:2014wwa} that in $E_6$

\beq
W_{E_6}=Tr\left(Y^3+X^4\right)
\eeq

\noi
or $E_8$

\beq
W_{E_8}=Tr\left(Y^3+X^5\right)
\eeq

\noi
dualities cannot be as simple as in the other cases. 

\section{\label{klasifikacija}Classification of the solutions}

Eq. (\ref{primerjava}) is our central equation. We will limit ourselves to the case of all charges 
$R_a\in[0,2]$. Although solutions of eq. (\ref{primerjava}) outside this domain exist, their interpretation 
is difficult. Not only, but this makes some of the powers negative and so the superconformal index cannot 
be expanded in powers of the fugacitiy $t$ atound $t=0$. This may even preclude the derivation of eq.  
(\ref{primerjava}).

With this assumption we can rewrite eq. (\ref{primerjava}) as 

\beq
\label{primerjava1}
\sum_{j=1}^\alpha t^{R_j}=\frac{t^{2\alpha\sum_{a=1}^{N_A+1}(R_a-1)}-1}
{\sum_{a=1}^{N_A+1}\left(t^{R_a}-t^{2-R_a}\right)}
\eeq

\noi
with $R_a\leq R_b$ if $a<b$ and $R_{N_A+1}=2$.

We will now further assume that all adjoints' $R$-charges are rational numbers, which is automatic if they 
are determined by a superpotential, and thus include a large family of cases. This will allow to make a change of 
variables:

\beq
\label{ty}
t^2=y^m
\eeq

\noi
with $m$ integer yet to be determined but large enough so that 

\beq
t^{2\alpha\sum_{a=1}^{N_A+1}\left(R_a-1\right)}-1=y^n-1
\eeq

\noi
with 

\beq
\label{nmalfa}
n=m\,\alpha\,\sum_{a=1}^{N_A+1}\left(R_a-1\right)
\eeq

\noi
integer. This can always be done since $R_a$ are rational numbers and $\alpha$ and $m$ (large enough) integers.

To classify all the solutions to eq. (\ref{primerjava1}), i.e. to find all integers $N_A$, $\alpha$, and 
rational numbers $R_a$, $a=1,\ldots,N_A$ and $R_j$, $j=1,\ldots,\alpha$, one essentially needs to 
factorise in all possible ways the polynomials $y^n-1$ for all integers $n>1$. We know that

\beq
\label{faktor}
y^n-1={\displaystyle \prod_{d|n}}\Phi_d(y)
\eeq

\noi
where $d$ run over all divisors of $n$ and $\Phi_n(y)$ is the 
{\it $n^{th}$ cyclotomic polynomial} \cite{cyclotomic}: for any positive  integer $n$, it is the unique 
polynomial with integer coefficients that is a divisor of $y^n-1$ and is not a divisor 
of $y^k-1$ for any $k<n$. Another way of defining it is as

\beq
\Phi_d(y)=\prod_{\substack{1\leq k\leq d\\gcd(k,d)=1}}\left(y-e^{2\pi ik/d}\right)
\eeq

\noi
where $gcd(k,d)$ is the greatest common divisor of $k$ and $d$. 
The degree of $\Phi_d(y)$ is equal to the 
{\it Euler's totient function} $\varphi(d)$.

The whole point is to rewrite the r.h.s. of eq. (\ref{faktor}) in all possible ways as

\beq
\label{produktplusminus}
{\displaystyle \prod_{d|n}}\Phi_d(y)=\Phi_n^+(y)\Phi_n^-(y)
\eeq

\noi
where $\Phi_n^+(y)$ is 
a polynomial with only non-negative (integer) coefficients, while $\Phi_n^-(y)$ is antipalindromic 
with integer coefficients\footnote{To prove it remember that all the cyclotomic polynomials are 
palindromic (like (\ref{antipalindromic}) but with $a_j = +a_{m-j}$) except $\Phi_1(y) = y-1$ which 
is antipalindromic. From the definition (\ref{produktplusminus}) we know that $\Phi_n^-(y)=(y^n-1)/\Phi_n^+(y)$, while $y^n-1$ 
contains $\Phi_1(y)$ exactly once. Since $\Phi_n^+(y)$ has only positive coefficients, 
it cannot contain $\Phi_1(y)$ and since the product of a palindromic polynomial with a (anti)palindromic polynomial 
is a (anti)palindromic polynomial, the result is that every $\Phi_n^-(y)$ is antipalindromic.}:

\beq
\label{antipalindromic}
\Phi_n^-(y)=\sum_{j=0}^ma_jy^j\;\;\;,\;\;\;a_j=-a_{m-j}
\eeq

As an example, take $n=6$. Then 

\beq
y^6-1=\Phi_1(y)\Phi_2(y)\Phi_3(y)\Phi_6(y)
\eeq

They can be found in appendix \ref{listciklo}: 

\bea
\Phi_1(y)&=&y-1\\
\Phi_2(y)&=&y+1\\
\Phi_3(y)&=&y^2+y+1\\
\Phi_6(y)&=&y^2-y+1
\eea

All the possibilities for $\Phi^+(y)$ and $\Phi^-(y)$ are shown in table \ref{nje6}. 

\begin{table}%[htdp]
\begin{center}
\resizebox{\textwidth}{!}{%
\begin{tabular}{|c|c|}
\hline
$\Phi_6^+(y)$ & $\Phi_6^-(y)$ \cr
\hline
\hline
$\Phi_2(y)=y+1$ & $\Phi_1(y)\Phi_3(y)\Phi_6(y)=y^5-y^4+y^3-y^2+y-1$ \cr
$\Phi_3(y)=y^2+y+1$ & $\Phi_1(y)\Phi_2(x)\Phi_6(x)=y^4-y^3+y-1$ \cr
$\Phi_2(y)\Phi_3(y)=y^3+2y^2+2y+1$ & $\Phi_1(y)\Phi_6(y)=y^3-2y^2+2y-1$ \cr
$\Phi_2(y)\Phi_6(y)=y^3+1$ & $\Phi_1(y)\Phi_3(y)=y^3-1$ \cr
$\Phi_3(y)\Phi_6(y)=y^4+y^2+1$ & $\Phi_1(y)\Phi_2(y)=y^2-1$ \cr
$\Phi_2(y)\Phi_3(y)\Phi_6(y)=y^5+y^4+y^3+y^2+y+1$ & $\Phi_1(y)=y-1$ \cr
\hline
\end{tabular}}
\end{center}
\caption{The possible products of cyclotomic polynomials with all coefficients positive $\Phi_n^+(y)$ 
and the corresponding antipalindromic polynomials $\Phi_n^-(y)=(y^n-1)/\Phi_n^+(y)$ for $n=6$.}
\label{nje6}
\end{table}%

In other words we rewrite eq. (\ref{faktor}) as

\beq
\label{faktor1}
\Phi_n^+(y)=\frac{y^n-1}{\Phi_n^-(y)}
\eeq

\noi
which it has exactly the form of eq. (\ref{primerjava1}). 

All we have to do is to find all factorisations (\ref{faktor1}), i.e. 
all subsets ${\cal D}^+_n\subset{\cal D}_n$ with 

\beq
{\cal D}_n=\{d;\;d|n\}
\eeq

\noi
for which the polynomial 

\beq
\label{Finplus}
\Phi^+_n(y)=\prod_{i\in{\cal D}^+_n}\phi_i(y)
\eeq

\noi
has only non-negative coefficients, so can be written as

\beq
\Phi^+_n(y)=\sum_{j=1}^{\alpha}y^{q_j}
\eeq

\noi
with $q_i\leq q_j$ for $i<j$.

Then we can rewrite

\beq
 \Phi^-_n(y)=\prod_{i\in{\cal D}^-_n}\phi_i(y)
\eeq

\noi
with

\beq
\label{Dplusminus}
{\cal D}^-_n\cup{\cal D}^+_n={\cal D}_n\;\;\;,\;\;\;{\cal D}^-_n\cap{\cal D}^+_n=\emptyset
\eeq

\noi
as

\beq
 \Phi^-_n(y)=\sum_{a=1}^{N_A+1}\left(y^{p_a}-y^{m-p_a}\right)
 \eeq

\noi
where $p_a\leq p_b$ for $a<b$ and 

\beq
m=p_{N_A+1}=\sum_{i\in{\cal D}^-_n}\varphi(i)
\eeq

\noi
with $\varphi(i)$ the $i^{\rm th}$ Euler's totient function, i.e. the highest power of the cyclotomic polynomial $\phi_i(y)$.
It is now obvious that 

\bea
\label{Ra}
R_a&=&\frac{2}{p_{N_A+1}}p_a\;\;\;,\;\;\;a=1,\ldots,N_A+1\\
\label{Rj}
R_j&=&\frac{2}{p_{N_A+1}}q_j\;\;\;,\;\;\;j=1,\ldots,\alpha
\eea

At this point we can calculate $\alpha$ from (\ref{nmalfa}):

\beq
\label{alfa}
\alpha=\frac{n}{\sum_{a=1}^{N_A+1}\left(2p_a-p_{N_A+1}\right)}
\eeq

In Appendix \ref{proof} we present a proof that $\alpha$ calculated from (\ref{alfa}) is always positive.

\subsection{Redundancy}

Although the classification of all solutions in the $y$-variable is on the one side useful because, since $n$ 
is integer, one cannot miss any solution (as long as it is not at too large $n$), it is on the other side redundant, 
since many different $n$ can give the same solution in the $t$-variable. 

As a simple example on what we have in mind, take $n=4$:

\beq
\label{yna4}
y^4-1=\Phi_1(y) \Phi_2(y) \Phi_4(y) 
\eeq

Then

\bea
\label{Fi4plus}
\Phi_4^+(y)&=&\Phi_2(y)=1+y\\
\label{Fi4minus}
\Phi_4^-(y)&=&\Phi_1(y)\Phi_4(y)=y^3-y^2+y-1
\eea

\noi
gives, according to (\ref{Ra}) and (\ref{Rj}), a solution $N_A=1$, $\alpha=2$

\bea
\label{Ran4}
R_a&=&\left(\frac{2}{3},2\right)\\
\label{Rjn4}
R_j&=&\left(0,\frac{2}{3}\right)
\eea

\noi
corresponding to eq. (\ref{primerjava}) of the form

\beq
\label{primerjavan4}
1+t^{2/3}=\frac{t^{8/3}-1}{t^2-1+t^{2/3}-t^{8/3}}
\eeq

\noi
which can be got directly from (\ref{yna4}), (\ref{Fi4plus}), (\ref{Fi4minus}) by using (\ref{ty}):

\beq
t^2=y^3
\eeq

A completely equal solution is for $n=8$, where

\beq
y^8-1=\Phi_1(y) \Phi_2(y) \Phi_4(y) \Phi_8(y) 
\eeq

\noi
if we choose

\bea
\Phi_8^+(y)&=&\Phi_4(y)=1+y^2\\
\Phi_8^-(y)&=&\Phi_1(y)\Phi_2(y)\Phi_8(y)=y^6-y^4+y^2-1
\eea

A different change of variable 

\beq
t^2=y^6
\eeq

\noi
gives the same equation (\ref{primerjavan4}) and thus the same solutions (\ref{Ran4}), (\ref{Rjn4}). 

A bit more general example is given in Appendix \ref{redundancy}.

There is however another type of redundancy. It is because if 

\beq
\Phi_n^+(y)=\frac{y^n-1}{\Phi_n^-(y)}
\eeq

\noi
is a solution, so is

\beq
\Phi_n^+(y)\sum_{m=1}^Ky^{n(m-1)}=\frac{y^{nK}-1}{\Phi_n^-(y)}
\eeq

\noi
since the left-hand-side is again a polynomial with only positive coefficients. The solution gives 
the same charges to the same number of adjoints ($\Phi_n^-(y)$, which determines them, does not change), 
but with a new number of mesons $K\alpha$ with charges

\beq
2R_Q+R_j+2(m-1)\alpha\sum_{a=1}^{N_A+1}\left(R_a-1\right)\;\;\;,\;\;\;j=1,\ldots,\alpha\;\;\;,\;\;\;m=1,\ldots,K
\eeq

This looks a bit puzzling: on one side the electric theory does not depend on the choice of $K$, so all global anomalies 
are independent on it as well. On the other side it is easy to show that 't Hooft anomaly matching conditions are 
satisfied as soon as the superconformal indices match. This means that in spite of all these new states, i.e. $(K-1)\alpha$ 
new mesons, nothing change in the anomalies of the magnetic dual. The reason is that the contribution of these 
new states are counterbalanced by the change of the magnetic colour, i.e. $\alpha N_f-N_c\to K\alpha N_f-N_c$ and 
with it the quark $R$-charge 

\beq
R_q\to1-\sum_{a=1}^{N_A+1}\left(R_a-1\right)\left(K\alpha-N_c/N_f\right)
\eeq

Before claiming that these new solutions represent new duals, one would need to do 
more checks. First, in some cases (as for example in SQCD) the $R$-charges of the 
new mesons are bigger than 2 and so outside the assumed interval. Second, finite 
$N_c$ could make these solutions disappear. However even if some of these solutions 
passed such tests, once we know the original solution ($K=1$), all the others ($K>1$) 
are easily got, so that we will not mention them (or count as new solutions) anymore 
in the following.

Let's now start scanning the solutions by increasing $n$. We divide them in increasing value of the number of adjoints $N_A$. 

\subsection{$N_A=0$ (SQCD)}

The simplest example is to take $\Phi^+(y)=1$. There is one such solution for 
each $n>0$:

\beq
1=\frac{y^n-1}{\prod_{i\in{\cal D}_n}\Phi_i(y)}
\eeq

All of them are just SQCD with only one meson $M\sim\tilde QQ$ in the magnetic theory.

\subsection{$N_A=1$}

The simplest non-trivial case is the one with only one adjoint. They can be classified by an integer $k>1$.

\subsubsection{\label{NAje1}$A_k$: $R_X=\frac{2}{k+1}$}

We get for $\alpha=k$ values of $R_j\equiv R(M_j)-2R_Q$:

\beq
\label{RjNA1}
R_j=\frac{2}{k+1}(j-1)\;\;\;,\;\;\;j=1,\ldots,k
\eeq

The first $15$ vales of $k$ are found from the first $30$ values of $n$ as can be seen from Table \ref{tabelaAk}, where 
only new solutions are shown. A given value $k$ appears for the first time for $n=2k$.

This is the well known case $A_k$ with the superpotential

\beq
W_{A_k}=Tr\left(X^{k+1}\right)
\eeq

\begin{table}%[htdp]
\begin{center}
%\resizebox{\textwidth}{!}{%
\begin{tabular}{|c|c|c||c|}
\hline
$n$ & ${\cal D}_n^+$ & ${\cal D}_n^-$ & $k$ \\
\hline
 4 & \{2\} & \{1,4\} & 2 \\
 6 & \{3\} & \{1,2,6\} & 3 \\
 8 & \{2,4\} & \{1,8\} & 4 \\
 10 & \{5\} & \{1,2,10\} & 5 \\
 12 & \{2,3,6\} & \{1,4,12\} & 6 \\
 14 & \{7\} & \{1,2,14\} & 7 \\
 16 & \{2,4,8\} & \{1,16\} & 8 \\
 18 & \{3,9\} & \{1,2,6,18\} & 9 \\
 20 & \{2,5,10\} & \{1,4,20\} & 10 \\
 22 & \{11\} & \{1,2,22\} & 11 \\
 24 & \{2,3,4,6,12\} & \{1,8,24\} & 12 \\
 26 & \{13\} & \{1,2,26\} & 13 \\
 28 & \{2,7,14\} & \{1,4,28\} & 14 \\
 30 & \{3,5,15\} & \{1,2,6,10,30\} & 15 \\
\hline
\end{tabular}
\end{center}
\caption{All solutions for $N_A=1$ and $n\leq30$ (first column) are defined by the set ${\cal D}_n^+$ 
(second column), from which one gets $\Phi_n^+(y)$ from (\ref{Finplus}). The third column 
is for completeness, and it is unique due to (\ref{Dplusminus}). The last column specifies the integer $k$ 
of the solution \ref{NAje1}.}
\label{tabelaAk}
\end{table}%

As an explicit example, let's see the case $n=12$:

\bea
\label{Fi12plus}
\Phi_{12}^+(y)&=&\Phi_2(y)\Phi_3(y)\Phi_6(y)=y^5+y^4+y^3+y^2+y+1\\
\Phi_{12}^-(y)&=&\Phi_1(y)\Phi_4(y)\Phi_{12}(y)=y^7-y^6+y-1
\eea

Since the powers of positive terms in $\Phi_{12}^-(y)$ are $(p_1,p_2)=(1,7)$, it follows that 

\beq
R_a=\frac{2}{7}\left\{1,7\right\}=\left\{\frac{2}{7},2\right\}
\eeq

\noi
and so $k=6$, as shown on Table \ref{tabelaAk}. The number of mesons are found from 
(\ref{alfa}), i.e. $\alpha=6$, confirmed by the $6$ terms in (\ref{Fi12plus}). Finally, the 
charges $R_j-2R_Q$ are found using (\ref{Rj}) and the powers $q_j$ in (\ref{Fi12plus}),

\beq
R_j=\frac{2}{7}(j-1)\;\;\;,\;\;\;j=1,\ldots,6
\eeq

\noi
confirming the $k=6$ case of (\ref{RjNA1}). 

We were searching for solutions for $N_A=1$ which cannot be cast into this form, but did not succeed 
in the limited range of finite $n$.

\subsection{$N_A=2$}

These are even more interesting examples. Up to $n=250$ we found 4 families of solutions, two already known, 
$D_{k+2}$ and $E_7$, a two new, one which we denote by $M_{k}$, and a mysterious one, which we denote by $N_{k}$. 
Let's now go through them.

\subsubsection{\label{Dk} $D_{k+2}$: $R_X=\frac{2}{k+1}$, $R_Y=\frac{k}{k+1}$}

The solution is for $\alpha=3k$ and 

\beq
R_j=\left(2(i-1)+k(m-1)\right)\frac{1}{k+1}\;\;,\;\;i=1,\ldots,k\;\;,\;\;m=1,2,3
\eeq

The superpotential is

\beq
W_{D_{k+2}}=Tr\left(X^{k+1}+XY^2\right)
\eeq

For $k$ odd one can get these results directly from the classical chiral ring. For $k$ even it is 
believed that this same conclusion follows from the quantum truncation of the chiral ring. 
However, in our approach we never use any reference to the superpotential, so there is 
no real difference between these two cases. Both satisfy equation (\ref{primerjava}), and this 
is all we need.

Some of these solutions found in our approach are shown in Table \ref{tabelaDk}.

\begin{table}%[htdp]
\begin{center}
%\resizebox{\textwidth}{!}{%
\begin{tabular}{|c|c|c||c|}
\hline
$n$ & ${\cal D}_n^+$ & ${\cal D}_n^-$ & $k$ \\
\hline
6 & \{2,3\} & \{1,6\} & 2 \\
 12 & \{2,3,4,6\} & \{1,12\} & 4 \\
 18 & \{3,6,9\} & \{1,2,18\} & 3 \\
 18 & \{2,3,6,9\} & \{1,18\} & 6 \\
 24 & \{2,3,4,6,8,12\} & \{1,24\} & 8 \\
 30 & \{3,5,10,15\} & \{1,2,6,30\} & 5 \\
 30 & \{2,3,5,10,15\} & \{1,6,30\} & 10 \\
\hline
\end{tabular}
\end{center}
\caption{
All solutions for $N_A=2$ and $n\leq30$ of the $D_{k+2}$ family. They are defined by $n$  (first column) and 
the set ${\cal D}_n^+$ (second column), from which one gets $\Phi_n^+(y)$ from (\ref{Finplus}). The third column 
is for completeness, and it is unique due to (\ref{Dplusminus}). The last column specifies the integer $k$ 
of the solution \ref{Dk}.}
\label{tabelaDk}
\end{table}%

\subsubsection{\label{E7}$E_7$}

This solution comes from $n=30$, which divisors are $1,2,3,5,6,10,15,30$. More precisely, 

\bea
\Phi_{30}^+(y)&=&\Phi_2(y)\Phi_3(y)\Phi_5(y)\Phi_6(y)\Phi_{10}(y)\Phi_{15}(y)\non\\
&=&y^{21}+y^{19}+y^{18}+y^{17}+2 y^{16}+y^{15}+2 y^{14}+2 y^{13}+2 y^{12}+2 y^{11}\non\\
&+&2 y^{10}+2 y^9+2 y^8+2 y^7+y^6+2 y^5+y^4+y^3+y^2+1\\
\Phi_{30}^-(y)&=&\Phi_1(y)\Phi_{30}(y)=y^9-y^7-y^6+y^3+y^2-1
\eea

Using the formulae of section \ref{klasifikacija} we find $N_A=2$, $\alpha=30$ and

\bea
R_a&=&\left\{\frac{4}{9},\frac{2}{3}\right\}\\
R_j&=&\left\{0,\frac{4}{9},\frac{2}{3},\frac{8}{9},\frac{10}{9},\frac{10}{9},\frac{4}{3},\frac{14}{9},\frac{14}{9},\frac{16}{9},
\frac{16}{9},2,2,\frac{20}{9},\frac{20}{9},\right.\non\\
&&\left.\frac{22}{9},\frac{22}{9},\frac{8}{3},\frac{8}{3},\frac{26}{9},\frac{26}{9},\frac{28}{9},\frac{28}{9},\frac{10}{3},\frac{32}{9},\frac{32}{9},\frac{34}{9},4,\frac{38}{9},\frac{14}{3}\right\}
\eea

\noi
which agrees with \cite{Kutasov:2014yqa}.

\subsubsection{\label{Mk}New solution, $M_k$: $R_X=\frac{2}{2k+1}$, $R_Y=\frac{2(k+1)}{2k+1}$}

This is a new possibility and it seems coming from the classical superpotential

\beq
\label{W2k}
W_{M_k}=Tr\left(X^{2k+1}+X^kY\right)
\eeq

It is easy to find that the solution with $\alpha=k$ and 

\beq
\label{RjMk}
R_j=(j-1)\frac{2}{2k+1}\;\;\;,\;\;\;j=1,\ldots,k
\eeq

We will see later on in section \ref{chiralring} why is the chiral ring given by (\ref{RjMk}).

Some lowest $k$ solutions are shown in table \ref{tabelaMk}. We did not include the case $k=1$, 
which is SQCD, so already part of the $N_A=0$ family.

\begin{table}%[htdp]
\begin{center}
%\resizebox{\textwidth}{!}{%
\begin{tabular}{|c|c|c||c|}
\hline
$n$ & ${\cal D}_n^+$ & ${\cal D}_n^-$ & $k$ \\
\hline
6 & \{2\} & \{1,3,6\} & 2 \\
 9 & \{3\} & \{1,9\} & 3 \\
 12 & \{2,4\} & \{1,3,6,12\} & 4 \\
 15 & \{5\} & \{1,3,15\} & 5 \\
 18 & \{2,3,6\} & \{1,9,18\} & 6 \\
 21 & \{7\} & \{1,3,21\} & 7 \\
 24 & \{2,4,8\} & \{1,3,6,12,24\} & 8 \\
 27 & \{3,9\} & \{1,27\} & 9 \\
 30 & \{2,5,10\} & \{1,3,6,15,30\} & 10 \\
\hline
\end{tabular}
\end{center}
\caption{
All solutions for $N_A=2$ and $n\leq30$ of the $M_{k}$ family. They are defined by $n$  (first column) and 
the set ${\cal D}_n^+$ (second column), from which one gets $\Phi_n^+(y)$ from (\ref{Finplus}). The third column 
is for completeness, and it is unique due to (\ref{Dplusminus}). The last column specifies the integer $k$ 
of the solution \ref{Mk}.}
\label{tabelaMk}
\end{table}%

We postpone to section \ref{interpretacijaMk} the interpretation of this result.

\subsubsection{\label{Nk}Apparent new solution, $N_k$: $R_X=\frac{6}{3+2k}$, $R_Y=\frac{2k}{3+2k}$}

This mysterious new solution has for the $\alpha=2k$ $R$-charges $R_j=R(M_j)-2R_Q$ 

\beq
R_j=\left(3(i-1)+k(m-1)\right)\frac{2}{3+2k}\;\;\;,\;\;\;i=1,\ldots,k\;\;,\;\;m=1,2
\eeq

Some of these solutions, for specific values of $k$, can be related to known solutions: 

\bea
\label{N1}
N_1&=&M_2\\
\label{N3k}
N_{3k}&=&D_{2k+2}
\eea

\noi
which is confirmed by some entries in table \ref{tabelaNk}, where the lowest $k$ solutions are shown. 

\begin{table}%[htdp]
\begin{center}
%\resizebox{\textwidth}{!}{%
\begin{tabular}{|c|c|c||c|}
\hline
$n$ & ${\cal D}_n^+$ & ${\cal D}_n^-$ & $k$ \\
\hline
 6 & \{2\} & \{1,3,6\} & 1 \\
 6 & \{2,3\} & \{1,6\} & 3 \\
 12 & \{2,4,6\} & \{1,3,12\} & 2 \\
 12 & \{2,3,4,6\} & \{1,12\} & 6 \\
 18 & \{2,3,6,9\} & \{1,18\} & 9 \\
 24 & \{2,4,6,8,12\} & \{1,3,24\} & 4 \\
 24 & \{2,3,4,6,8,12\} & \{1,24\} & 12 \\
 30 & \{2,5,10,15\} & \{1,3,6,30\} & 5 \\
 30 & \{2,3,5,10,15\} & \{1,6,30\} & 15 \\
\hline
\end{tabular}
\end{center}
\caption{
All solutions for $N_A=2$ and $n\leq30$ of the $N_{k}$ family. They are defined by $n$  (first column) and 
the set ${\cal D}_n^+$ (second column), from which one gets $\Phi_n^+(y)$ from (\ref{Finplus}). The third column 
is for completeness, and it is unique due to (\ref{Dplusminus}). The last column specify the integer $k$ 
of the solution \ref{Nk}.}
\label{tabelaNk}
\end{table}%

With a generic $k$, the only marginal superpotential we can write with these $R$-charges is 

\beq
\label{WNk}
W_{N_k}=Tr\left(XY^2\right)
\eeq

\noi
the main obstacle being non-integer powers needed to sum up to $R=2$. This is the $D$ theory in the 
classification \cite{Intriligator:2003mi}, where $a$-maximisation \cite{Intriligator:2003jj} is needed to determine 
all charges. This is now a problem: although the $R$-charges for the adjoints guarantee that the expressions 
for the electric and magnetic superconformal indices, and so formally the $a$-central charges, are the same, 
the $a$ central charges are {\it not} maximised, in spite of the fact that one $R$-charge is not determined by the 
superpotential constraints. While one could naively think that it is possible to obtain the needed  
rational adjoints' $R$-charges by the maximisation procedure for discrete choices of $x=N_c/N_f$, this is possible 
only for one among the electric and magnetic theories, not both. This is not surprising, two equations 
$\partial a^{el}/\partial R_Y=0$ and $\partial a^{mag}/\partial R_{\tilde Y}=0$ in general cannot be satisfied by the same 
choice of $x$. At the moment we cannot offer any interpretation of this solution except in special cases $k=1$ (\ref{N1}) 
or $k$ multiple of 3 (\ref{N3k}).

\section{\label{nova}More on the new solution \ref{Mk}}

We will now first comment on the possible interpretation of the new solution found in \ref{Mk} and 
then see how its quantum chiral ring looks like.

\subsection{\label{interpretacijaMk}Possible interpretation}

How to understand the new solutions $M_k$ from section \ref{Mk}, especially in view of the classification of 
\cite{Intriligator:2003mi}? We propose that it could be a low energy limit of the $D_{2k+2}$ theory

\beq
\label{alltermsD}
W_{D_{2k+2}}=Tr\,\left(X^{2k+1}+XY^2\right)
\eeq

Adding the relevant operator

\beq
\label{extra}
\Delta W=Tr\,\left(X^kY\right)
\eeq

\noi
two different things can happen:

\bet

\item
in the low energy theory the first term of (\ref{alltermsD}) dominates giving our new solution

\beq
W_{M_k}=Tr\,\left(X^{2k+1}+X^kY\right)
\eeq

\begin{figure}[h] 
   \centering
   \includegraphics[width=4.in]{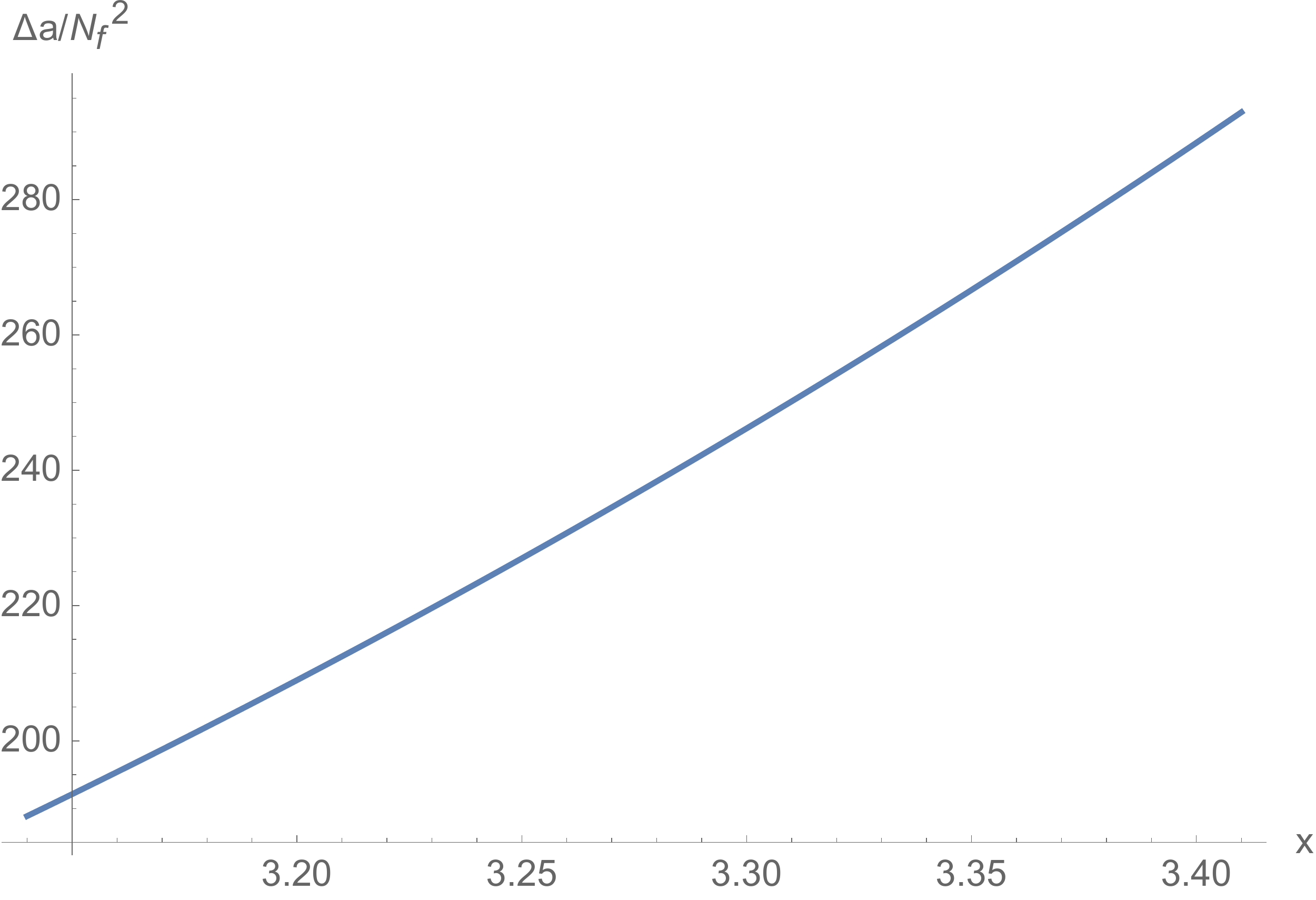} 
   \caption{The difference of the $a$ central charges between the supposedly UV fixed point $D_6$ and the supposedly IR 
   fixed point $M_2$.}
   \label{deltaaD2kplus2}
\end{figure}

Take as an example $k=2$. Then $x_{D_6}^{min}\approx3.14$ \cite{Intriligator:2003mi}, $a_{M_2}$ and $c_{M_2}$ stay 
positive as they should \cite{Anselmi:1997ys}, while from around $x_{M_2}^{max}\approx3.41$ on the collider bound \cite{Hofman:2008ar} 
starts being violated. But in the interval $x_{D_6}^{min}\leq x\leq x_{M_2}^{max}$ the difference $\Delta a=a_{D_6}-a_{M_2}$ 
is positive and thus satisfies the $a$-theorem \cite{Zamolodchikov:1986gt,Cardy:1988cwa,Osborn:1989td,Jack:1990eb,
Komargodski:2011vj,Komargodski:2011xv}, as shown in fig.\ref{deltaaD2kplus2};

\item
it is the second term of (\ref{alltermsD}) which dominates over the first one:

\beq
\label{druga}
W=Tr\,\left(XY^2+X^kY\right)
\eeq

This case gives the $R$-charges of the adjoints the same as in $D_{2k}$\footnote{We could have added a term 
$X^{k+1}Y$ also in (\ref{alltermsD}), since it is allowed by $R$-symmetry in this case of $D_{n+2}$ 
for even $n=2k$, which makes the fixed point $D_{2k+2}$ actually a fixed line \cite{Leigh:1995ep,Strassler:2005qs}.}. 
By assuming that (\ref{druga}) has also a dual in the present classification through the cyclotomic polynomials, 
than this {\it is} the theory $D_{2k}$. 

\eet

In short, the new candidate for a fixed point $M_k$ can be a low energy limit of $D_{2k+2}$ with 
a perturbation (\ref{extra}) added, presumably for a large enough ratio of the Yukawas $y_{X^{2k+1}}/y_{XY^2}$.

\subsection{\label{chiralring}The chiral ring}

It is strange that all the mesons we need for the duality of the 
case in section \ref{Mk} are just $\alpha=k$, i.e. $\tilde QX^{j-1}Q$ for $j=1,\ldots,k$. In fact, the superpotential is

\beq
W=Tr\left(X^{2k+1}+X^kY\right)
\eeq

The e.o.m. for this system are

\bea
\label{XYplusYX}
\frac{\partial W}{\partial X}&=&X^{2k}+X^{k-1}Y+X^{k-2}YX+\ldots+XYX^{k-2}+YX^{k-1}\propto I\\
\frac{\partial W}{\partial Y}&=&X^k\propto I
\eea

\noi
from which it is not clear among others why there is no $Y$ in the mesons. 

As an example we will now consider the case $k=2$. Following \cite{Cachazo:2001gh,Intriligator:2003mi} 
we can easily see that classically at most single powers of $X$ and $Y$ are independent, while (\ref{XYplusYX}) 
for $k=2$ tells us that $X$ and $Y$ anticommute (modulo identity operator). So the possible mesons are naively

\beq
\tilde QX^{i-1}Y^{m-1}Q\;\;\;,\;\;\;i=1,2\;\;\;,\;\;\;m=1,2
\eeq

But this is only apparent, which we will see now, following the suggestion of 
\cite{Kutasov:2014wwa}, through the superconformal index. To keep track of the gauge invariant terms we 
use an expansion of (\ref{gaugeinvariantindex}) with the following values for the electric theory\footnote{
We could have done the same for the magnetic theory. The expansion in gauge invariant operators would 
obviously look differently, but the equality of the two superconformal indices would guarantee that the 
number of them in short multiplets is the same and that there is a one-to-one matching between the two sets 
of independent operators.}

\begin{align}
\label{fn}
f_E(w^n)&=\frac{(Ft^2)^n+(d_+\lambda_-t^2)^n-(t\lambda_+)^n-(t\lambda_-)^n+
\sum_{a=1}^{N_A}\left((X_at^{R_a})^n-(\bar{\psi}_{X_a}t^{2-R_a})^n\right)}{\left(1-(td_+)^n\right)\left(1-(td_-)^n\right)}\\
\label{gn}
g_E(w^n)&=\frac{(Qt^{R_Q})^np_{N_f}(y^n)-(\bar{\psi}_{\tilde Q}t^{2-R_Q})^np_{N_f}(\tilde y^n)}
{\left(1-(td_+)^n\right)\left(1-(td_-)^n\right)}v^n\\
\label{gbarn}
\bar g_E(w^n)&=\frac{(\tilde Qt^{R_Q})^np_{N_f}(\tilde y^{-n})-(\bar{\psi}_{Q}t^{2-R_Q})^np_{N_f}(y^{-n})}
{\left(1-(td_+)^n\right)\left(1-(td_-)^n\right)}v^{-n}
\end{align}
 
\noi
where we denoted by $\bar{\psi}_{X,Y}$ the anti-fermions of the adjoints $X,Y$, by $d_{+,-}$ the derivatives, by 
$\lambda_{+,-}$ the gauginos (with only one combination among $d_+\lambda_-$ and $d_-\lambda_+$ independent 
due to the equation of motion \cite{Dolan:2008qi,Kinney:2005ej}) and $F$ the gauge field strength. 

Eqs. (\ref{fn}), (\ref{gn}) and (\ref{gbarn}) have been derived from (\ref{fE}), (\ref{gE}) and (\ref{gEbar}) by explicitly 
denoting the origin of each single term (the fugacity $x$ ($x^{-1}$) has been replaced by $d_+$ ($d_-$) or 
$\lambda_+$ ($\lambda_-$)), following for example \cite{Dolan:2008qi}, see also table 2 of \cite{Kinney:2005ej} and 
table 1 of \cite{Gadde:2010en} for a list of elements which contribute to the counting.

Now we can expand (\ref{gaugeinvariantindex}) as

\beq
{\cal I}_E(w)={\cal I}_E^{(0)}(t)+{\cal I}_E^{(1)}(t)\left(\tilde QQ\,t^{2R_Q}p_{N_f}(y)p_{N_f}(\tilde y^{-1})+\ldots\right)+\ldots
\eeq

First we see the gauge invariant operators without the quark 
fields by expanding in powers of $t$:

\bea
\label{IE0}
{\cal I}_E^{(0)}&=&1+t^{4/5} X^2+t^{6/5} \left(X^3-X \bar{\psi }_Y\right)-\left(\lambda _-+\lambda _+\right) t^{7/5} X+t^{8/5} X \left(-X \bar{\psi }_Y
+2 X^3+Y\right)\non\\
&+&t^{9/5} \left(\lambda _- \left(\bar{\psi }_Y-X^2\right)+\lambda _+ \bar{\psi }_Y+d_- X^2+d_+ X^2-\lambda _+ X^2\right)\non\\
&+&t^2 \left(-2 X^3 \bar{\psi }_Y-X \bar{\psi }_X+X \bar{\psi }_Y^2-Y \bar{\psi }_Y+\lambda _- \lambda _++2 X^5+X^2 Y\right)+O\left(t^{11/5}\right)
\eea

\noi
while the index of the terms proportional to $\tilde QQ$ is

\bea
\label{IE1}
{\cal I}_E^{(1)}&=&1+t^{2/5} X+t^{4/5} \left(2 X^2-\bar{\psi }_Y\right)+t \left(2 d_-+2 d_+-\lambda _--\lambda _+\right)+t^{6/5} \left(-3 X \bar{\psi }_Y
+3 X^3+Y\right)\non\\
&+&3 t^{7/5} X \left(d_-+d_+-\lambda _--\lambda _+\right)+t^{8/5} \left(-6 X^2 \bar{\psi }_Y-\bar{\psi }_X+\bar{\psi }_Y^2+5 X^4+3 X Y\right)\non\\
&+&t^{9/5} \left(d_- \left(7 X^2-3 \bar{\psi }_Y\right)-3 d_+ \bar{\psi }_Y+\lambda _- \left(3 \bar{\psi }_Y-6 X^2\right)+3 \lambda _+ \bar{\psi }_Y
+7 d_+ X^2-6 \lambda _+ X^2\right)\non\\
&+&t^2 \left(-11 X^3 \bar{\psi }_Y-3 X \bar{\psi }_X+5 X \bar{\psi }_Y^2-3 Y \bar{\psi }_Y+d_- \left(4 d_+-3 \lambda _--3 \lambda _+\right)
-3 d_+ \lambda _+\right.\non\\
&+&\left.\lambda _- \left(3 \lambda _+-2 d_+\right)+3 d_-^2+3 d_+^2+F+\lambda _-^2+\lambda _+^2+7 X^5+6 X^2 Y\right)+O\left(t^{11/5}\right)
\eea

The first term in (\ref{IE1}), i.e. $1$, means the operator $\tilde QQ$. The second term, $X$, represents the operator $\tilde QXQ$. 
The third one, $2X^2-\bar{\psi}_Y$, means that there is the operator $\tilde QQ\,Tr(X^2)$, while $\tilde QX^2Q$ gets paired 
with $\tilde Q\bar{\psi}_YQ$ forming a long multiplet and thus escaping the counting of the superconformal index, which is sensible 
only to short multiplets \cite{Romelsberger:2005eg}.

One can continue, finding that at each level only gauge invariant operators made out of $\tilde QQ$ or $\tilde QXQ$ 
multiplied by gauge invariant operators without quarks allowed by (\ref{IE0}). 
For example, let's see why there is no $\tilde QYQ$ in the counting. This is a term of order $t^{6/5}$ in (\ref{IE1}). The number of all 
gauge invariant operators coming from short multiplets of this order is $1$, distributed as

\bea
\label{t65minus3}
-3&:&\tilde QX\bar{\psi}_YQ\;,\;\tilde Q\bar{\psi}_YXQ\;,\;\tilde QQ\,Tr(X\bar{\psi}_Y)\\
\label{t65plus3}
+3&:&\tilde QX^3Q\;,\;\tilde QXQ\,Tr(X^2)\;,\;\tilde QQ\,Tr(X^3)\\
+1&:&\tilde QYQ
\eea

We know from the term of order $t^{6/5}$ of (\ref{IE0}) that the last term of (\ref{t65minus3}) pairs with the last term 
of (\ref{t65plus3}). Thus the only multi trace term remained and allowed by (\ref{IE0}) is $\tilde QXQ\,Tr(X^2)$, 
so there is no room for $\tilde QYQ$ to survive unpaired. 

The reader can continue in this fashion: the absence of $\tilde QXYQ$ as a short multiplet is simply because 
the only 2 possibilities are 

\beq
\tilde QQ\,\left(Tr(X^2)\right)^2\;,\;\tilde QQ\,Tr(XY)
\eeq

\noi 
which are products of already allowed terms.

In other words, the only single trace operators made out of one quark $Q$ 
and one antiquark $\tilde Q$ are exactly $\tilde QQ$ and $\tilde QXQ$, and no others, in accord with the meson 
operators with $R$-charge $2R_Q+R_j$ needed in the magnetic theory. So the naive expectation from the classical 
equations of motion for the mesons present is misleading. Quantum constraints take care of this, which can be explicitly 
seen in the expansion of the superconformal index, as suggested in \cite{Kutasov:2014wwa} and presented here for 
the case of interest.

\section{\label{conclusion}Conclusions}

Which are the possible fixed points in SQCD theories with $N_A$ ($\geq0$) adjoints? A list of them, all connected to a UV free 
theory, can be found in \cite{Seiberg:1994pq,Kutasov:2003iy,Intriligator:2003mi}. All of them except the cases denoted by 
$E_6$ and $E_8$ \cite{Kutasov:2014wwa} have a known candidate for a dual theory. In this work we gave a complete 
parametrisation of all such theories with a prescribed dual, assuming the adjoints' $R$-charges are rational numbers in the interval 
between $0$ and $2$. These solutions include all known cases and give new ones. A family of new such solutions have 
been given, as well as possible suggestions on where these new solutions could lay on the tree of flows.

It is not clear from the classification whether there exist different classes of solutions to those presented in this work. The reason is that 
at least in principle one should go through an ordered but infinite number of possibilities, and check whether they are new or 
already in the known families. The mathematical problem boils down to possible classification of all products of distinct 
cyclotomic polynomials with positive coefficients. We do not know a solution.

In this work we limited the adjoints' charges to rational numbers in the interval between $0$ and $2$. The choice of the 
rational numbers is mandatory for the classification of the solutions through the factorisation of the polynomial $y^n-1$ 
into products of cyclotomic polynomials, while the choice of the interval $[0,2]$ is less obvious\footnote{Considerations of unitarity 
bounds \cite{Mack:1975je} could help here, although we were unable to conclude either way. In fact gauge non-invariant fields could in 
principle have $R$-charges below $2/3$ or even negative, as long as all gauge invariant combinations  satisfy the unitarity bound.}. 
The motivation for 
it has two reasons.

First, in many cases it might be difficult to find a superpotential\footnote{Without a superpotential the $R$-charges must be 
defined through the $a$-maximisation \cite{Intriligator:2003jj}, which typically gives non-rational solutions, see section \ref{Nk} .} 
that enforces negative $R$-charges. For example, the solution of section \ref{NAje1} can be formally enlarged to $k<-1$. 
However we do not know of a superpotential with positive powers of the fields which would enforce it, unless one uses extra 
singlets\footnote{If this is done in both the electric and magnetic versions, their effect cancels out in the difference 
of the superconformal index.}. 

A second, and, in our opinion, more important reason is that it is difficult to define a sensible superconformal index. Due to 
negative powers of the fugacity $t$ the superconformal index does not have a Taylor expansion in powers of 
$t$ around the origin. This, among others, means that even the derivation of eq. (\ref{primerjava}) is not guaranteed. We plan 
to look at this problem better in future, although it may well be that such $R$-charges outside the domain $[0,2]$ are 
simply forbidden.

In this work we limited ourselves to theories with at most two adjoints ($N_A\leq 2$). Of course it is straighforward to 
find solutions for general $N_A$, and we did it. The problem is that these theories are not asymptotically free and 
so, if connected to the free theory, typically violate the $a$-theorem. The only exceptions found so far are theories with 
at least some adjoints' $R$-charges negative. These could represent examples of UV dual fixed points, i.e. examples of 
UV safety \cite{Litim:2014uca} in supersymmetric theories \cite{Intriligator:2015xxa,Bajc:2016efj,Bond:2017suy,Bajc:2017xwx}. 
Due to the difficult interpretation of these cases we leave also this analysis for the future.

Last but not least, all the analysis of this paper is done in the leading large $N_c$ limit. 
Only in this case the equality of the electric and magnetic superconformal indices 
reduces to a simple and easily calculable expression. So even the original equation 
(\ref{primerjava}) or equivalently (\ref{primerjava1}), from which we started the analysis, is not known in 
general for finite $N_c$ and to get it requires much more effort, which is beyond 
the scope of this paper. On one side it is in principle possible that the new solutions 
$M_k$ (and/or $N_k$) are just an artifact of the large $N_c$ expansion and disappear 
when finite number of colours are considered. On the other side we are optimistic 
since the known solutions $D_{k+2}$ and $E_7$ persist for finite $N_c$, although 
clearly there is no guarantee that this is true also for the new candidate $M_k$.

\subsubsection*{Acknowledgments}
We acknowledge the financial support from the Slovenian Research Agency (research core funding No.~P1-0035). 
We thank Steve Abel and Francesco Sannino for discussion on various issues considered in this paper.

\begin{appendices}
%\appendixpageoff

\section{\label{listciklo}A list of cyclotomic polynomials}

For convenience we explicitly present here all the cyclotomic polynomials up to $n=30$. This choice of the 
upper $n$ is not dictated by what one can find in \cite{cyclotomic}, but by the value needed for $E_7$ in section \ref{E7}.

{\allowdisplaybreaks
\begin{align}
\Phi_{1}(y)&= y-1 \non\\
\Phi_{2}(y)&= y+1 \non\\
\Phi_{3}(y)&= y^2+y+1 \non\\
\Phi_{4}(y)&= y^2+1 \non\\
\Phi_{5}(y)&= y^4+y^3+y^2+y+1 \non\\
\Phi_{6}(y)&= y^2-y+1 \non\\
\Phi_{7}(y)&= y^6+y^5+y^4+y^3+y^2+y+1 \non\\
\Phi_{8}(y)&= y^4+1 \non\\
\Phi_{9}(y)&= y^6+y^3+1 \non\\
\Phi_{10}(y)&= y^4-y^3+y^2-y+1 \non\\
\Phi_{11}(y)&=   y^{10}+y^9+y^8+y^7+y^6+y^5+y^4+y^3+
   y^2+y+1 \non\\
\Phi_{12}(y)&= y^4-y^2+1 \non\\
\Phi_{13}(y)&=   y^{12}+y^{11}+y^{10}+y^9+y^8+y^7+y^
   6+y^5+y^4+y^3+y^2+y+1 \non\\
\Phi_{14}(y)&= y^6-y^5+y^4-y^3+y^2-y+1 \non\\
\Phi_{15}(y)&= y^8-y^7+y^5-y^4+y^3-y+1 \non\\
\Phi_{16}(y)&= y^8+1 \non\\
\Phi_{17}(y)&=  \sum_{i=1}^{17} y^{i-1} \non\\
\Phi_{18}(y)&= y^6-y^3+1 \non\\
\Phi_{19}(y)&= \sum_{i=1}^{19} y^{i-1} \non\\
\Phi_{20}(y)&= y^8-y^6+y^4-y^2+1 \non\\
\Phi_{21}(y)&= y^{12}-y^{11}+y^9-y^8+y^6-y^4+y^3-y+1\non\\
\Phi_{22}(y)&=   y^{10}-y^9+y^8-y^7+y^6-y^5+y^4-y^3+
   y^2-y+1 \non\\
\Phi_{23}(y)&= \sum_{i=1}^{23} y^{i-1} \non\\
\Phi_{24}(y)&= y^8-y^4+1 \non\\
\Phi_{25}(y)&= y^{20}+y^{15}+y^{10}+y^5+1 \non\\
\Phi_{26}(y)&=   y^{12}-y^{11}+y^{10}-y^9+y^8-y^7+y^
   6-y^5+y^4-y^3+y^2-y+1 \non\\
\Phi_{27}(y)&= y^{18}+y^9+1 \non\\
\Phi_{28}(y)&= y^{12}-y^{10}+y^8-y^6+y^4-y^2+1 \non\\
\Phi_{29}(y)&=  \sum_{i=1}^{29} y^{i-1} \non\\
\Phi_{30}(y)&= y^8+y^7-y^5-y^4-y^3+y+1 
\end{align}}

\section{\label{proof}A proof that $\alpha>0$ and a sum rule}

One could worry that the quantity $\alpha$ defined in (\ref{alfa}) may not be 
positive, and thus invalidate the consistency. Here we show that it is always positive, as it must be. It follows 
from the general property of the cyclotomic polynomials for $n>1$ ($\Phi_1(1)=0$):

\bea
\label{cyclotomicat1}
\Phi_n(1)&=&1\;\;{\rm if}\;n\;{\rm is\; not\; a\; prime\; power}\\
\Phi_n(1)&=&p\;\;{\rm if}\;n=p^k\;{\rm is\; a\; prime\; power\; with}\;k\geq1\non
\eea

This means that any product of cyclotomic polynomials with $n\ne1$ is positive at $y=1$. 
In our case 

\beq
\label{aty0}
\left.\frac{\Phi_n^-(y)}{y-1}\right|_{y\to1}>0
\eeq

But we can expand

\bea
\frac{\Phi_n^-(y)}{y-1}&=&\frac{1}{y-1}\sum_{a=1}^{N_A+1}\left(y^{p_a}-y^{m-p_a}\right)\non\\
&=&\frac{1}{y-1}\sum_{a=1}^{N_A+1}\left(\Theta(2p_a-m)y^{m-p_a}\left(y^{2p_a-m}-1\right)
-\Theta(m-2p_a)y^{p_a}\left(y^{m-2p_a}-1\right)\right)\non\\
&=&\sum_{a=1}^{N_A+1}\left(\Theta(2p_a-m)\sum_{j=0}^{2p_a-m-1}y^{m-p_a+j}-
\Theta(m-2p_a)\sum_{j=0}^{m-2p_a-1}y^{p_a+j}\right)
\eea

Taking it at $y=1$ we get

\bea
\left.\frac{\Phi_n^-(y)}{y-1}\right|_{y\to1}&=&\sum_{a=1}^{N_A+1}\left(\Theta(2p_a-m)(2p_a-m)-
\Theta(m-2p_a)(m-2p_a)\right)\non\\
&=&\sum_{a=1}^{N_A+1}\left(2p_a-m\right)
\eea

\noi
which in combination with (\ref{aty0}) gives 

\beq
\label{pam}
\sum_{a=1}^{N_A+1}\left(2p_a-m\right)>0
\eeq

\noi
Eq. (\ref{alfa}) (remember that $m=p_{N_A+1}$) finally proves that $\alpha>0$.

From here it immediately follows a sum rule for $\sum_{a=1}^{N_A+1}\left(R_a-1\right)$. 
Since 

\beq
\sum_{a=1}^{N_A+1}\left(R_a-1\right)=\frac{1}{p_{N_A+1}}\sum_{a=1}^{N_A+1}\left(2p_a-p_{N_A+1}\right)
\eeq

\noi
due to (\ref{pam}) we also find that

\beq
\label{sumrule}
\sum_{a=1}^{N_A+1}\left(R_a-1\right)>0
\eeq

This gives a necessary (although not suffficient) criterium which any given set of adjoints' $R$-charges, which represent 
a Seiberg-Kutasov dual theory, must satisfy. For example the choice $R_a=(1/6,1/3)$ cannot. In fact for this case 
$\sum_{a=1}^{3}=(1/6-1)+(1/3-1)+(2-1)=-1/2<0$ and thus violates the sum rule \ref{sumrule}.

\section{\label{redundancy}Redundancy}

Let $n$ be the product of two prime numbers $p_{1,2}$ and an integer $r>0$:

\beq
n=rp_1p_2
\eeq

Then  choose

\beq
\label{finminus}
\Phi_{n}^-(y)=\Phi_1(y^r)\Phi_{p_1p_2}(y^r)
\eeq

The divisors of $p_1p_2$ are $1$, $p_1$, $p_2$ and $p_1p_2$, so

\beq
\Phi_{n}^+(y)=\frac{y^n-1}{\Phi^-(y)}=\Phi_{p_1}(y^r)\Phi_{p_2}(y^r)
\eeq

\noi
has all coefficients positive, since for any prime $p$

\beq
\Phi_p(x)=\sum_{i=1}^px^{(i-1)}
\eeq

With

\beq
y^{r(1+\varphi(p_1p_2))}=t^2
\eeq

\noi
equation (\ref{primerjava}) now becomes

\beq	
\Phi_1\left(t^{\frac{2}{1+\varphi(p_1p_2)}}\right)\Phi_{p_1p_2}\left(t^{\frac{2}{1+\varphi(p_1p_2)}}\right)=
\frac{t^{\frac{2p_1p_2}{1+\varphi(p_1p_2)}}-1}
{\Phi_{p_1}\left(t^{\frac{2}{1+\varphi(p_1p_2)}}\right)\Phi_{p_2}\left(t^{\frac{2}{1+\varphi(p_1p_2)}}\right)}
\eeq

\noi
with $\varphi(n)$ the $n^{th}$ Euler's totient function. Once $p_{1,2}$ are fixed, solution (\ref{finminus}) 
does not lead to new solutions for different $r$. Of course, the ansatz (\ref{finminus}) is only one possibility, and others 
are possible, among them for example ${\cal D}_n^-=\{1,p_1,p_1p_2\}$ or ${\cal D}_n^-=\{1,p_2,p_1p_2\}$.

The main message is that increasing $n$ does not necessarily give new solutions. This might 
mean that the number of families is finite and thus in some future possible to determine all of them.

\end{appendices}

\end{document}